Elevated Tc levels in YBa2Cu3O6.5 modeled on a 3-dimensional doped bond structure of chain and plane pairs

H. Oesterreicher, Department of Chemistry, UCSD, La Jolla, CA  92093-0506

The complex phenomenology of shot quenched YBa2Cu3O6.5 with Tc=100K and 200K levels is compared with laser pulsed analogs with an eye on explaining the presumed Tc=552K of the latter. Shot quenching can produce metastable states with pronounced increases in plane metric and cell volume, accompanied by a rough doubling of Tc to a 100K level of an orthorhombic with 3-fold O coordinated chain Cu (O3 type). These states decay over a non-superconducting transition range to the conventional Tc=50K level of O24. We consider the plane expanded laser pulsed materials to contain aspects of O42 plane n-doped counterparts of the O3 n-doped version of shot quench preparations. In addition, we assume that highly charged p-doped chains of 4-fold O coordination form hole pairs at trijugate position, allowing close approach of the apical O to the electron-doped planes. They are now capable of participating in the bonding with the plane pairs at corresponding $3a_0/2$ location. The overall pair number is therefore multiplied, and the coupling strengthened, by limited 3-D effects within the Plane-Chain-Plane sandwich. The latter can be seen as an extended chemical bonding system that has the potential to equilibrate contractive and expansive pairs and so obviate the need for distinction of doping type as it may exchange it dynamically. It is argued that indications for a Tc=200K level on shot quenching has a related origin and represents one in series of predicted Tc levels based on bond order principles. Predictions are made where similar effects can be expected in other compound classes.

Introduction
Short infrared laser pulses have recently been shown [1] to drive a conventionally prepared YBa2Cu3Oy=6.5 ceramic with Tc=52K to one showing coherent interlayer transport strongly reminiscent of superconductivity up to Tc=552K. Accompanying the electronic transition is a structural one: the 5-fold O coordinated copper dioxide double layers (2)5 become more separated by two picometers as well as more buckled, and the layer distance between them becomes thinner by a similar amount. We shall refer to this layer, comprised by the chain system and the sandwiching 2 planes, as the PCP sandwich. Also da, the distance of Cu to the apical O decreases. It was suggested that laser pulses cause individual atoms in the crystal lattice to shift briefly and thus enhance the superconductivity.

Here we connect these findings with studies on influencing self-doping through introducing internal tension by preparative variety such as shot quenching near O half filling of YBa2Cu2CuOy [2-6], written here in a way to separate the plane and chain sites. Changing lattice pressure with temperature influences the competition between plane and chain sites, especially where the latter can occur in various local environments. It can create exotic metastable materials with various forms of tension self-doping including reversals of doping types and indications of Tc enhancements hinting at the achievability of strong 3D-coupling through pairs on chains at a Tc=200K level. While the conventionally prepared materials show contracted plane and cell volume (designated V-), there exist plane and cell volume expanded (designated V+) as well as plane expanded-strongly c-axis contracted (C-) phases with varying symmetries, axial ratios and superconducting properties on shot quenching.

As the V+ types have been explained by a change in doping to electron type on planes, we will similarly introduce the notion that laser pulsing causes related effects, combined with the creation of additional pairs on chains.

Generally, systems with apical O are not tending to be electron doped on planes. However, this exclusionary principle can be relaxed in high-energy environments, which tend to require more space for planes. In a further conceptual step for laser pulsing we consider that the self-doping on planes, of expanding electron type, can draw the apical O close enough to the planes for direct bonding. This allows the establishment of a balancing bond order of holes on chains that can be transformed into pairs and introduce 3-D coupling within the PCP sandwich. In this sense the deleterious effect of the apical O will have been turned to advantage as this usually weak bond with the plane is now transformed into a hole doped pair bond on either side of the chain sticks. In fact, if a direct 3-D propagation of these expansive n-doped and contractive p-doped pairs occurs, then an averaging of their effects would be expected, so that from a phase stability point these materials would resemble aspects of the semiconductor they are derived from.

It appears that the mercurial O half filling situation of $YBa_2Cu_3O_{6.5}$ can display a variety of Tc levels. Conventionally prepared, materials have Tc~50K but shot quenching raises this to 100K (with indications for 200K, or 0K in the transitional range) and laser pulsing results now in a Tc~550K level. We will therefore draw on the changeable nature of the local environment at O half filling and the variety of self doping levels in bond order structures, both on planes and chains, for explaining the corresponding Tc level structure. The general nature of this bond or charge ordering in electronic crystals has been seen in STM and ARPES [7,8] where indications for pairs at $3a_0/2$ location were observed in several charge-lattice commensurate bond order patterns on planes. Such pairs at $3a_0/2$ should in principle also be a possibility on chains and can be expected to order themselves into the plane-plaid bond order.

Generally, an algorithm relates the approach of the apical O in its deleterious effect on Tc, potentially extinguishing it [4-6]. We will here indicate that this algorithm can revert on even closer approach of the apical O. We will show that the chains have several requisites for bond ordering similar to the one observed in the planes. This involves creating pairs in trijugate positions (doped bonds separated by 2 intervening un-doped bonds or a distance of $3a_0/2$). In one modification the hole-doped bond would lie between the apical O and the plane Cu on the two ends of the chain sticks. These pairs would be multiplying the number of overall pairs. This postulated pair creation on chains would then allow for aspects of strongly 3-dimensionally correlated superconductivity within the PCP sandwich, by contrast with the usual assumption of ceramic superconductivity being mainly confined to the planes, and chain approach having a deleterious effect on Tc.

Results

In order to appreciate the uniqueness of the laser pulsed materials' Tc we first embed them within a phenomenological approach. Generally, a relation exists [5,6] between Tc and da as witnessed in the Hg analogs compared to the Cu ones. We follow an empirical formalism Tc=ic where c is the charge concentration and i=fTe is a measure of the plane's isolation. Empirically Te=600K for hole doped materials and f is a function of da through a formalism involving bond valence. f~2/3 for some materials with one apical O approaching the plane as in the present case, so that i=400K. For conventional V- with c=0.125 one calculates Tc= 400x0.125=50K, in line with observation. On searching for a theoretical limit in Tc we look at the case of lack of hybridization between plane and apical O with f=1, and we set c=0.25 as the presently obtainable optimum charge concentration. Accordingly one calculates Tc=150K, as obtained in the Hg analogs. The reported value of Tc=552K therefore lies strongly outside common phenomenology and with c=0.125 would demand a recalibrated i in the form of Tc=Ic with I=4416K for super-high Tc. We propose therefore that origins of strong superconductivity enhancement lie in aspects of novel strong 3-dimensional correlations. In fact, the short distance to the apical O and the short distance between pairs of planes (PCP sandwich) could be a hint for an unusual beneficial interaction out of the planes. To illuminate on this point we will now introduce the local redox machinery.

Generally, $YBa_2Cu_2CuO_{6.5}$ is composed of a twofold Cu plane site that is 5-fold O-coordinated, denoted (2)5. The one fold chain site can occur as 3-fold O-coordinated (1)3, which is a component in the corresponding semiconductor at lower y. $YBa_2Cu_2CuO_{6.5}$ is special in that it is formally charge balanced with plane and chain sites at 2+. However, when obtained conventionally by slow cooling to below 300K, it obtains as an orthorhombic, predominantly composed of alternate Cu chains of 2-fold O coordination (1)2 and 4-fold O coordination (1)4, denoted O24 or ortho II, with (1)2 and (1)4 formally at 1+ and 3+ respectively. Yet, it shows relative plane contraction, indicative of contracting hole or p-doping on planes and volume expanding n-doping on (1)4 in an internally self-doped way. Cell volume also contracts to 173A3 (designated V- type) by comparison with the tetragonal semiconductor at lower y, extrapolated to y=6.5 (174.5A3 of cell-volume neutral or V0 type). This can be seen as a result of self-doping, where hole placement acts as a phase stability adjuster. In particular, it can be understood as (2)5 being oxidized to 2.125+ and (1)4 being reduced to 2.5+ while (1)2 stays at 1+. This phenomenology corresponds to the conventional p-doped V- type. It has been reasoned that the series of c=0.125=2/16, c=0.167=2/12 and c=0.222=2/9 corresponds to pair repeat patterns on planes of 4a0x4b0, 3a0x4b0 and 3a0x3b0 respectively, all found in STM or ARPES [7,8] and manifesting themselves in $YBa_2Cu_2CuO_y$ in the various well known Tc plateaus with y. Extra stability is gained through charge-lattice commensurability. Several data support this critical charge concentration scheme. For the related materials at higher y=6.67 this is known to create holes to c=0.17 on planes according to Knight shift [9], supporting the Tc=66K plateau on some preparation. For y=6.92 Knight shift shows c=0.22, supporting the 90K plateau. For conventionally prepared $YBa_2Cu_3O_{6.5}$ we extrapolate then c=0.125. This c=0.125 corresponds [5,6] to a special electronic crystal with the plaid pair repeat distances 4a0 and 4b0 [7,8]. This charge has to balance with chains so that c on (2)5 has to equal 4c on (1)4 from stoichiometry. For (2)5 with 2.125+, one calculates then charge on (1)4 as 2.5+. On shot quenching, several high temperature modifications can be stabilized at room temperature. In order to slow phase transitions we have used $YBa_2Cu_{3-x}Ni_xO_y$ with minute x<0.05 [2]. These modifications can be connected with variation in chain coordination assigned to calculations based on O-O potentials [10]. They represent steps within the equilibrium (1)2 + (1)4=2(1)3 that plays some role in all preparations. An interesting modification within the present context is a more or less isotropic cell volume expanded (177.5A3 corresponding to 2.6% expansion relative to O24 and designated V+ type) orthorhombic variety near y=6.46 that shows (1)3 in TEM. This O3, of V+ type, has increased bulk Tc=72K, compared with O24 V- type, even though Tc is generally reduced by the partial Ni substitution (extrapolated Tc=99K without Ni, representing a Tc=100K level). This plane-expanded state was explained as a result of electron or n-doping on planes.

We assume that the optimum doping of this O3 of V+ type happens at c=0.222, corresponding theoretically to y=6.44 (a small amount of (1)2 must be left at this composition). This special c=0.222 requirement has to do with the stability, through charge-lattice commensurability, of a pair bond order with repeat distances of 3a0x3b0 [5,6]. In fact, optimum Tc for O4 is experimentally found around y=6.9, just where one would expect the (1)4 dopant to provide a similar bond order with 3a0x3b0 and corresponding c=0.222 (it also harbors therefore residual amounts of (1)2), corresponding to a theoretical y=6.889. Both situations therefore involve a comparable bond ordering on planes, albeit of different doping type and dopant origin, bearing out the charge-lattice commensurability arguments. However, while y=7 has overshot the c=0.222 bond order for conventional preparations, y=6.5 lies at the right composition for generating c=0.125 corresponding to bond ordering with 4a0x4b0 of laser pulsed materials.

There exists further variety in the shot quench phenomenology potentially germane to the laser doped materials. One state, centering around y=6.7 with 174A3 and denoted C-, displays severe c-axis contraction and large plane expansion leading to a pseudo-cubic condition with lattice parameters a~b~c/3=3.87A. It is non-superconducting and of V0 type. We assign it to a plane n-doped derivative of O3 of T34 parentage where some 1(3) are transformed to 1(4) due to stoichiometry. We assume its theoretical stoichiometry to be y=6.667, where there would be c=0.167 from 1(3) and c= 0.0833 from 1(4), making for a total c=0.25=4/4x4. C- reminds of a Pr analog at y near 7 that can be prepared with a deleterious small da. It appears, though, to represent aspects of the state of laser pulsed materials with their strongly reduced distance to the apical O and assumed strong coupling between planes and chains. We speculate that it contains localized pairs on chains and planes, which could perhaps become de-localized and resemble the laser pulsed phenomenology when prepared closer to y=6.5 where critical conditions for charge order commensurability between planes and chains are met. Even some tetragonal materials at y<6.5 have been prepared with strong cell volume expansions and the Tc=50K level, making them interesting for probing with laser pulsing for 3-D effects.

Generally, some shot quenched preparations can show diamagnetism to 150K (extrapolated Tc=203K without Ni) corresponding to small sample components. They could represent a further step in gaining properties, characteristic of laser pulsing, perhaps incurred in the transition of O3 to O24. We note that no direct explanation is left for a further doubling of Tc to a 200K level solely on the above charge ordering arguments and we will have to imply a novel mechanism. As with the stronger effects on Tc of laser pulsing we will also in this case invoke the involvement of pairs on chains further below under discussion.

Summarizing the experimental situation one notes for shot quenching, at the highest cell volume expansion, an O3 type with small amounts of a Tc=200K component and a majority phase at level Tc=100K. On ageing at room temperature, this material shrinks, going through a non-superconducting state, only to have superconductivity reappear with a conventional Tc=50K level in the conventional O24 structure, which has now only half the active doping component. For the laser pulsed materials, structure appears to stay at O42 with Tc=550K. This state decays in an unknown way at room temperature to the Tc=50K level. We shall now attempt to discuss these levels in terms of variation in self-doping and a new source of pairs on chains.

Discussion

High-energy environments, such as elevated temperature or laser pulsing, lead to accentuated plane expansions for YBa2Cu3O6.5. At elevated temperature this is accompanied by a shift in equilibrium (1)2 + (1)4=2(1)3 to the right. When these materials are rapidly quenched to room temperature, a new type of tension self-doping can occur. Accordingly a crucial assignment for the combination of cell volume expansion and elevated Tc levels for these plane-expanded shot quenched materials is a flip in doping from p-type on planes to n-type. This is augmented by principles of enhanced doping through the rough doubling of dopant (1)3 of O3 compared to (1)4 in O24 simply by stoichiometry. This naturally explains the rough doubling of Tc from the Tc=50K level to the 100K level from O24 to O3 as outlined above (the accurate relation should be 0.222/0.125 and be involving isolation factor i). We will now start with ideas on the Tc=552K level that have to involve new principles. With those principles we can then explore the response to altered conditions leading potentially to even higher Tc or explain lower Tc levels such as the 200K one. While we do not have definitive information on the crucial parameters such as pair numbers on various sites, we can attempt to indirectly extract them from phenomenology, and profitably delineate aspects of a story in the genesis of unusual behavior.

In principle one could assume that the effects on laser pulsing also originate from a shift in the equilibrium (1)2 + (1)4=2(1)3 towards the right. However, time scales of atomic rearrangement make this less likely by comparison with purely electronic rebalancing. For electronic rebalancing we could assume that strong plane expanding vibrations, induced by laser pulsing, tend to reduce the charge on plane Cu to an n-doped state, while chains become p-doped. As this alone would hardly suffice for the strength of the observed phenomena we look for more unique situations growing out of this situation, involving the third dimension. In particular, we consider situations in which the electronic crystal corresponding to the doping on planes would be commensurate with the projection of the doped chain system perpendicular to it which would also be transformed into pairs. This 3-D situation would comprise the full PCP sandwich and would lend strength to the superconductivity. We shall now consider the origins of these interactions by exploring the relationship of the O3 of V+ type to laser pulsed O24 analogs.

Following the method for calculating charge or bond ordering in conventionally prepared O24 material with y=6.5 and c=0.125 we note a bond order on planes with charge repeat distances $4a_0$ and $4b_0$ as this allows to contain one doped pair according to 2/4x4=0.125. The conventional O24 preparation therefore has (2)5 at 2.125+ and (1)4 at 2.5+ presumably in the form of a doped bond from Cu to O as a single. If one now assumes the plane expansion of laser pulsed materials to indicate a flip of doping to n-type on planes then formally speaking its self-doped charge on (2)5 is symmetrically disposed at 1.875+. This corresponds to charge on (1)4 of 3.5+ while we take (1)2 to stay inert at 1+, although it could in principle also share in the doping.

We note now that for the laser pulsed O42 the unusually high charges on (1)4 of 3.5+ should allow close approach of the apical O to the negatively doped planes in the sense of a jump of the doped bond to the location between chain O and plane Cu. This is assumed to induce direct participation in charge transport and bonding, involving a charge pattern where alternating sticks have doped bonds between apical O and plane Cu instead of on the chain Cu side of the apical O. It is then possible that positive charge is further organized to the opposite ends of these doped chain sticks and so creates correlated hole pairs on (1)4 at the equivalent of $3a_0/2$ or trijugate position perpendicular to the plane. Every fourth stick could then accommodate a pair, leading naturally to commensurability with the plane pair bond order of 4a0x4b0 type. Accordingly, on adding these pairs to the plane pairs, the overall pair concentration has now doubled from c=0.125=2np to 2np =0.25 but the short da would, at face value, not allow superconductivity. Instead, we expect that the short da introduces new 3-D coupling, strengthening superconductivity in a major way, and so explain the Tc=552K. Calibrating to this value we can calculate Tc=Ic with I=4416K and c corresponding to the value characteristic of the planes (alternatively one defines Tc=4416np, where np is the total pair number). We want to indicate that an alternative location for pairs on chains could be along the O, connecting the chain Cu. However, this would not reconcile the expected contractions, on hole doping into this bond location, with the conditions imposed by plane expansion. It is also more difficult to understand strong 3-D effects, although bonding loops with the planes would be possible.

At this point it is interesting to speculate on the limits of effects. Taking c=0.222, we calculate Tc=981K. This would correspond to the 3-D coupled counterpart of the O3 material at Tc=100K level. We note, however, that the fortuitous situation of 4a0x4b0, in which pairs on planes within 2a0 leave an equal 2a0 space open for contact with pairs on chains, could rule out 3-D effects at higher c. This restriction would not apply for lower c, where bond order with 5a0x5b0 can occur at c=0.080, and one calculates Tc=353K on chain pair formation. Such a situation may occur in the transition of O3 to O42 after shot quenching, if we assume that newly generated (1)4 carries the pairs. For the observed Tc=200K one calculates c=0.045, which usually corresponds to about the onset condition for superconductivity which may, in fact, now occur at considerably lower c when strong 3-D effects are involved. It could reflect the plane bond order 6a0x6b0 with c=0.0556. Similar to 4a0x4b0, it allows symmetric placement of plane and chain pairs, which is not the case with 5a0x5b0. We therefore tentatively assign the Tc=200K level to this origin.

Following the idea that superconductivity in 3-D trijugate systems could occur at lower c compared to 2-D situations, bond order with 8a0x8b0 and c=0.031 could result in Tc near 137K. Bond order with 4a0x5b0 and c=0.1 accordingly would have Tc=442K. These situations could occur in O24 at y=6.4 either on laser pulsing or through special preparation. In this connection it would be interesting to further monitor the decay of O3 to O24 concerning a string of possible Tc levels of transitory products. Some could perhaps even be observed in situ during elevated temperature synthesis given their unusual strength. A similarly interesting project would be to monitor the region in y corresponding to onset of orthorhombic splitting and superconductivity, as here various amounts of (1)4 are generated in a narrow range. It is, however, possible that temperature induced lattice pressure is a required condition for stability of 3-D trijugate systems. In this case situations have to be induced that provide internal tensions. Novel preparative methods hold continuing interest in this respect.

It remains an open question whether (1)3 can in principle similarly be induced to show pairs on chains. The diminished charge on (1)3, by comparison with (1)4 of O24, may not be able to penetrate directly to the planes. As outlined above, the one case where 3-D effects may already have been observed, outside laser pulsing, could be the Tc=200K level in preparations with bulk O3 of Tc=100K level where (1)4 could be responsible for this minority component. In this sense we would already have an example at hand, for a stabilized material at the Tc=200K level, displaying the operation of strong 3-D effects in cuprate superconductors. However, we have not answered to the question of whether (1)3 can in principle support 3-D superconductivity. Assuming that it can, one can propose a conventionally prepared material with y=6.25 for laser pulsing. This could induce self-doping to c=0.125 and the next question would be if 3-D effects develop simultaneously; in other words, is Tc of the 50K or of the 550K level.

We do presently not know details of the mechanism of charge transport in cuprate superconductors and can therefore only conjecture on mechanism within their presumed 3-D analog plaid patterns of the canonical 2-D counterpart. We can assume that positive and negative pairs will search each other out but that phonons will keep them apart and in motion. This may involve exchanging their position from plane to chain. In fact, we note that the formal charge on (1)4 has already lost some of its meaning as the doped chain pair is now closer to the plane Cu. It would further loose meaning on pair exchange. We note that an averaging of doping types, in the process of charge transport, would alleviate the strong anisotropies or volume expansion effects. In fact, such averaging would roughly correspond to the phase stability criteria of the non-superconducting semiconductor from which the conventional superconductor is derived. This taking recourse to the stability criteria of the parent material may in fact be reflected in the considerably smaller cell volume expansion effects, estimated at 0.9% on laser pulsing, compared with the 2.6% on V+ formation. The laser pulsed material is therefore close to designation as V0.

In this context, we consider now the contracted and strongly correlated PCP sandwich system as a new chemical motive. It represents a 3-D network of coupled trijugate bonds. By comparison with the bijugate system, say in benzene, charge is arranged with properties of a localized electronic crystal, and so not fully smeared out. We assume that the chain pairs are commensurably nestled with the plane bonds along the lines of the plaid pattern. The coupling between adjacent planes, of the 3-dimensionally active sandwich structure, can then be assumed to be more indirect and not primarily responsible for the high Tc. We note that the C- phase also displays the essential features of plane expansion and thin PCP sandwich of laser pulsed materials. However, it stays electronically localized, presumably due to incommensurable charge ordering and disorder on the chains through crowding. It may though represent a precursor to the laser pulsed superconductors and be transformable into a room temperature superconductor if its stability can be extended to y=6.5.

Given the mercurial behavior, near O half filling of chains, one can now consider to probe, with laser pulsing, some of the unique modifications of $YBa_2Cu_{3-x}Ni_xO_{6.5}$ obtained on shot quenching. The relative blocking of structural transitions, through increasing x, could result in achieving relative stabilization of room temperature Tc. In the light of laser pulsed phenomenology more sophisticated rapid quenching techniques may also achieve, per se, relatively stable materials with Tc>300K. We assume that it may also be possible to extend 3-D phenomenology to other compound classes with an eye on those that exhibit short da (usually highly charged chain equivalents) with the potential to flip to yet shorter ones, exhibiting the respective geometry such as single layer Tl analogs.

Generally, the present ideas rest on the assumption that high Tc superconductivity manifests itself as a bond order phenomenon with tendencies to a harmonic Tc level structure. This Tc level structure reflects the period of charge ordering and the degree of isolation of the planes but as conjectured earlier, it can also include their connectivity out of the planes. The computational tools and theoretical framework need now to be developed for understanding how multiple influences work together to produce overall charge ordering and transport. What is now needed is an encompassing crystal chemistry of high Tc superconductivity, or more specifically, the energetic of charge placement as a phase stability question. Open questions are many, including: can 3-D effects obtain at c>0.125 or with different local environments and other structure types. Quantum chemical definition of the proposed

5-center bond for the pair and its 3-D interaction needs to be explored and its network architecture elaborated on. The mechanism of charge transport needs clarification.

In closing we note that the projects of inducing novel behavior through the counter play of local self-doping, achievable through preparative finesse, has been quite successful for some time and has recently received a welcome boost in the form of results from laser pulsing.

Summary

Indications for unusually high temperature superconductivity are explained in terms of the creation of additional pairs on chains connected with a flip to a dynamically alternating doping type, resulting in strong 3-D correlations. Accordingly, the laser pulsed materials are considered to represent aspects of an $O_{42}$ n-doped counterpart, of $4a_0 \times 4b_0$ type bond ordering, to the O3 plane expanded n-doped version of the shot quenched materials of $3a_0 \times 3b_0$ type. The highly charged p-doped chains then allow close approach of the apical O to the planes, participating directly in the bonding. This can create pairs also within the chains that can now couple with the planes within the 3-D Plane-Chain-Plane sandwich describable by $T_c=4416c$. The concept of a 3-D pair coupled PCP sandwich can be seen as an extended chemical unit of coupled trijugate bonds that can obviate the need for distinction of local doping type as it may exchange it dynamically. Several aspects in the phenomenology attest to the activity of charge-lattice commensurability of a pair bond order at $3a_0/2$ location including $O_{24}$, $O3$. As another example the minority phase with $T_c=200K$ level of shot quenched materials is similarly discussed as based on a transitory $O_{324}$ phase of $6a_0 \times 6b_0$ type with 3-D interactions. For C- type phases we assume localized pairs on chains because of bond order constraints at $4a_0 \times 3b_0$. However, by extending the phase to lower $c=0.125$ it may become a $T_c>300K$ material. Although the suitability of (1)3 for 3-D effects is an open question, one could probe for it in a tetragonal $y=6.25$ material with laser pulsing as it has the self-doping potential to $c=0.125$.